\documentclass[12pt]{article}

\usepackage[margin=1in]{geometry}
\usepackage{amsmath,amssymb,mathrsfs,amsthm}
\newcommand{\model}{\text{model}}

\usepackage{natbib}
\usepackage{booktabs, array, ragged2e}
\usepackage{float}
\usepackage{rotating}
\usepackage{subcaption}
\usepackage[normalem]{ulem}
\usepackage[dvipsnames]{xcolor}

\usepackage{tikz}
\usetikzlibrary{shapes,shadows,arrows,shapes.misc}
\tikzstyle{latent}=[rounded corners, draw=black,
 text centered, node distance=4em, text width=4em]
\tikzstyle{latent_circle}=[draw=black, circle, minimum size=4em,
 text centered, node distance=12em, text width=3em]
 \tikzstyle{observed}=[ draw=black,
 text width=1.4em, text centered, node distance=4em,]
 \tikzstyle{observed_square}=[ draw=black,  minimum size=5.66em,
 regular polygon,
 regular polygon sides=4,
 text centered, 
 node distance=12em,]
\tikzstyle{line}=[draw,-latex']
\tikzstyle{dline}=[draw,dashed,-latex']
\tikzstyle{nline}=[draw,strike out,-latex']

\begin{document}

\title{Identifiability of the Multinomial Processing Tree-IRT model for the Philadelphia Naming Test \\ Technical Report}

%
%
%

\author{Andrew Womack$^1$ \and Daniel Taylor Rodriguez$^2$ \and Gerasimos Fergadiotis$^3$ \and William D. Hula$^{4,5,6}$}
\date{%
   \footnotesize{$^1$Rice University, Department of Statistics\\%
   $^2$ Portland State University, Department of Mathematics and Statistics\\%
   $^3$Portland State University, Department of Speech and Hearing Sciences\\%
   $^4$Geriatric Research, Education, and Clinical Center, VA Pittsburgh Healthcare System\\%
   $^5$Audiology and Speech Pathology Service, VA Pittsburgh Healthcare System\\%
   $^6$Department of Communication Science and Disorders, University of Pittsburgh} \\[2ex]%
  \normalsize{\today}
}

\maketitle

\section{Abstract}
Naming tests represent an essential tool in gauging the severity of aphasia and monitoring the trajectory of recovery for individuals afflicted with this debilitating condition. In these assessments, patients are presented with images corresponding to common nouns, and their responses are evaluated for accuracy. The Philadelphia Naming Test (PNT) stands as a paragon in this domain, offering nuanced insights into the type of errors made in responses. In a groundbreaking advancement, Walker et al. (2018) introduced a model rooted in Item Response Theory and multinomial processing trees (IRT-MPT). This innovative approach sought to unravel the intricate mechanisms underlying the various errors patients make when responding to an item, seeking to pinpoint the specific stage of word production where a patient's capability falters.  However, given the sophisticated nature of the IRT-MPT model proposed by Walker et al. (2018), it is imperative to scrutinize both its conceptual as well as its statistical validity. Our endeavor here is to closely examine the model's formulation to ensure its parameters are identifiable as a first step in evaluating its validity.

\section{Understanding the Problem}

\begin{figure}[h!]
\center
\tiny{
\begin{tikzpicture}
\node[latent](Att){Attempt};
\node[latent,right of=Att,xshift=6em](Sem){Sem};
\node[latent,right of=Sem,xshift=6em](LexSem){LexSem};
\node[latent,right of=LexSem,xshift=6em](LexPhon){LexPhon};
\node[latent,right of=LexPhon,xshift=11em](LexSel){LexSel};
\node[observed,below of=Att,](NA){NA};
\node[latent,below of=Sem](Phon1){Phon};
\node[latent,below of=LexSem](Phon2){Phon};
\node[latent,below of=LexPhon](Phon3){Phon};
\node[latent,below of=LexSel,xshift=-5em](Phon4){Phon};
\node[latent,below of=LexSel,xshift=5em](Phon5){Phon};
\node[latent,below of=Phon1,xshift=-2.5em](WordL1){Word-L};
\node[observed,below of=Phon1,xshift=2.5em](U1){U};
\node[latent,below of=Phon2,xshift=-2.5em](WordL2){Word-L};
\node[observed,below of=Phon2,xshift=2.5em](S){S};
\node[latent,below of=Phon3,xshift=-2.5em](WordT1){Word-T};
\node[observed,below of=Phon3,xshift=2.5em](F1){F};
\node[latent,below of=Phon4,xshift=-2.5em](WordT2){Word-T};
\node[observed,below of=Phon4,xshift=2.5em](M){M};
\node[latent,below of=Phon5,xshift=-2.5em](WordT3){Word-T};
\node[observed,below of=Phon5,xshift=2.5em](C){C};
\node[observed,below of=WordL1,xshift=-2em](AN1){AN};
\node[observed,below of=WordL1,xshift=2em](U2){U};
\node[observed,below of=WordL2,xshift=-2em](AN2){AN};
\node[observed,below of=WordL2,xshift=2em](U3){U};
\node[observed,below of=WordT1,xshift=-2em](N1){N};
\node[observed,below of=WordT1,xshift=2em](F2){F};
\node[observed,below of=WordT2,xshift=-2em](N2){N};
\node[observed,below of=WordT2,xshift=2em](F3){F};
\node[observed,below of=WordT3,xshift=-2em](N3){N};
\node[observed,below of=WordT3,xshift=2em](F4){F};
\path[line](Att)--(NA) node[pos=0.5,left]{$1-\psi_{1tk}$};
\path[line](Att)--(Sem) node[pos=0.5,above]{$\psi_{1tk}$};
\path[line](Sem)--(Phon1) node[pos=0.5,left]{$1-\psi_{2t}$};
\path[line](Sem)--(LexSem) node[pos=0.5,above]{$\psi_{2t}$};
\path[line](LexSem)--(Phon2) node[pos=0.5,left]{$1-\psi_{3tk}$};
\path[line](LexSem)--(LexPhon) node[pos=0.5,above]{$\psi_{3tk}$};
\path[line](LexPhon)--(Phon3) node[pos=0.5,left]{$1-\psi_{4tk}$};
\path[line](LexPhon)--(LexSel) node[pos=0.5,above]{$\psi_{4tk}$};
\path[line](LexSel)--(Phon4) node[pos=0.5,left]{$1-\psi_{5tk}$};
\path[line](LexSel)--(Phon5) node[pos=0.5,right]{$\psi_{5tk}$};
\path[line](Phon1)--(WordL1) node[pos=0.5,left]{$1-\psi_{6tk}$};
\path[line](Phon1)--(U1) node[pos=0.5,right]{$\psi_{6tk}$};
\path[line](Phon2)--(WordL2) node[pos=0.5,left]{$1-\psi_{6tk}$};
\path[line](Phon2)--(S) node[pos=0.5,right]{$\psi_{6tk}$};
\path[line](Phon3)--(WordT1) node[pos=0.5,left]{$1-\psi_{6tk}$};
\path[line](Phon3)--(F1) node[pos=0.5,right]{$\psi_{6tk}$};
\path[line](Phon4)--(WordT2) node[pos=0.5,left]{$1-\psi_{6tk}$};
\path[line](Phon4)--(M) node[pos=0.5,right]{$\psi_{6tk}$};
\path[line](Phon5)--(WordT3) node[pos=0.5,left]{$1-\psi_{6tk}$};
\path[line](Phon5)--(C) node[pos=0.5,right]{$\psi_{6tk}$};
\path[line](WordL1)--(AN1) node[pos=0.5,left]{$1-\psi_{8}$};
\path[line](WordL1)--(U2) node[pos=0.5,right]{$\psi_{8}$};
\path[line](WordL2)--(AN2) node[pos=0.5,left]{$1-\psi_{8}$};
\path[line](WordL2)--(U3) node[pos=0.5,right]{$\psi_{8}$};
\path[line](WordT1)--(N1) node[pos=0.5,left]{$1-\psi_{7k}$};
\path[line](WordT1)--(F2) node[pos=0.5,right]{$\psi_{7k}$};
\path[line](WordT2)--(N2) node[pos=0.5,left]{$1-\psi_{7k}$};
\path[line](WordT2)--(F3) node[pos=0.5,right]{$\psi_{7k}$};
\path[line](WordT3)--(N3) node[pos=0.5,left]{$1-\psi_{7k}$};
\path[line](WordT3)--(F4) node[pos=0.5,right]{$\psi_{7k}$};
\end{tikzpicture}
}
\caption{\label{fig:irtmpt1}
Original IRT-MPT model graph from Walker. Observable states have sharp corners and latent states have rounded corners.}
\end{figure}
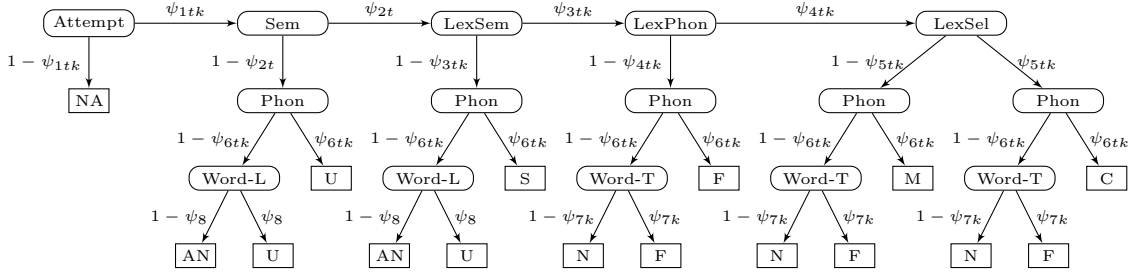

Naming tests represent an essential tool in gauging the severity of aphasia and monitoring the trajectory of recovery for individuals afflicted with this debilitating condition. In these assessments, patients are presented with images corresponding to common nouns, and their responses are meticulously evaluated for accuracy. The Philadelphia Naming Test (PNT) stands as a paragon in this domain, offering nuanced insights into the type of errors made in responses. In a groundbreaking advancement, \citep{walker} introduced a model rooted in Item Response Theory and multinomial processing trees (IRT-MPT). This innovative approach sought to unravel the intricate mechanisms underlying the various errors patients make when responding to an item. By acknowledging the heterogeneity of both items and individuals, this model aims to identify the specific stage of word production where a patient's capability falters by estimating multiple latent parameters for patients to more precisely determine at which step of word of production a patient's ability has been affected. These latent parameters aspire to provide a representation of the theoretical cognitive steps taken in responding to an item, as shown in Figure \ref{fig:irtmpt1}. Given the complexity of the model proposed in \citep{walker}, here we investigate the identifiability/estimability of the parameters included in the model referenced above.

The probabilities for an edge traversal to an observable node in Figure \ref{fig:irtmpt1} can either be global, local to respondent, local to item, or local to both. These probabilities arise from IRT models and so usually take the form of some link function evaluated at the difference between respondent ability and item difficulty. There are redundancies in the probabilities that lead to the model only using eight probabilities to describe the sixteen path probabilities. The model is completed with the specification of a distribution on respondent ability, item difficulty, or global probabilities. These distributions all take the form of a fixed distribution without any hyperparameters. The specification of the model proposed by \citep{walker} is that of a fixed effects model given that there is no sharing of information across parameters within or between individuals or items. The prior imposes regularization, which renders all parameters estimable in the sense that a proper posterior distribution is achieved. However, this does not mean that the parameters are identifiable. There are eight parameters and eight categorical observables, which suggests an over-parameterized and unidentifiable model.

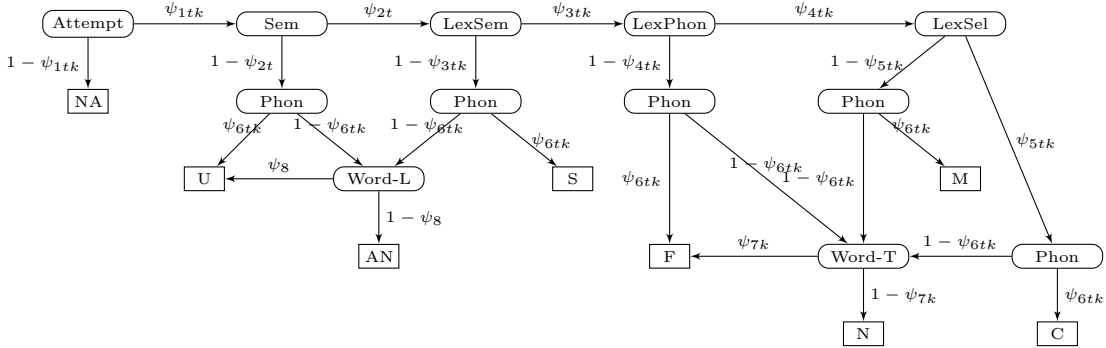
\begin{figure}[h!]
\center
\tiny{
\begin{tikzpicture}
\node[latent](Att){Attempt};
\node[latent,right of=Att,xshift=6em](Sem){Sem};
\node[latent,right of=Sem,xshift=6em](LexSem){LexSem};
\node[latent,right of=LexSem,xshift=6em](LexPhon){LexPhon};
\node[latent,right of=LexPhon,xshift=11em](LexSel){LexSel};
\node[observed,below of=Att,](NA){NA};
\node[latent,below of=Sem](Phon1){Phon};
\node[latent,below of=LexSem](Phon2){Phon};
\node[latent,below of=LexPhon](Phon3){Phon};
\node[latent,below of=LexSel,xshift=-5em](Phon4){Phon};
\node[latent,below of=LexSel,xshift=5em,yshift=-8em](Phon5){Phon};
\node[latent,below of=Phon1,xshift=5em](WordL){Word-L};
\node[observed,below of=Phon2,xshift=5em](S){S};
\node[observed,below of=Phon4,xshift=5em](M){M};
\node[latent,below of=Phon4,yshift=-4em](WordT){Word-T};
\node[observed,below of=Phon5](C){C};
\node[observed,below of=WordL](AN){AN};
\node[observed,left of=WordL,xshift=-5em](U){U};
\node[observed,below of=WordT](N){N};
\node[observed,left of=WordT,xshift=-6em](F){F};
\path[line](Att)--(NA) node[pos=0.5,left]{$1-\psi_{1tk}$};
\path[line](Att)--(Sem) node[pos=0.5,above]{$\psi_{1tk}$};
\path[line](Sem)--(Phon1) node[pos=0.5,left]{$1-\psi_{2t}$};
\path[line](Sem)--(LexSem) node[pos=0.5,above]{$\psi_{2t}$};
\path[line](LexSem)--(Phon2) node[pos=0.5,left]{$1-\psi_{3tk}$};
\path[line](LexSem)--(LexPhon) node[pos=0.5,above]{$\psi_{3tk}$};
\path[line](LexPhon)--(Phon3) node[pos=0.5,left]{$1-\psi_{4tk}$};
\path[line](LexPhon)--(LexSel) node[pos=0.5,above]{$\psi_{4tk}$};
\path[line](LexSel)--(Phon4) node[pos=0.5,left]{$1-\psi_{5tk}$};
\path[line](LexSel)--(Phon5) node[pos=0.5,right]{$\psi_{5tk}$};
\path[line](Phon1)--(WordL) node[pos=0.5,right,above]{$1-\psi_{6tk}$};
\path[line](Phon1)--(U) node[pos=0.5,left,above]{$\psi_{6tk}$};
\path[line](Phon2)--(WordL) node[pos=0.5,left,above]{$1-\psi_{6tk}$};
\path[line](Phon2)--(S) node[pos=0.5,right]{$\psi_{6tk}$};
\path[line](Phon3)--(WordT) node[pos=0.5,right,above]{$1-\psi_{6tk}$};
\path[line](Phon3)--(F) node[pos=0.5,left]{$\psi_{6tk}$};
\path[line](Phon4)--(WordT) node[pos=0.5,left]{$1-\psi_{6tk}$};
\path[line](Phon4)--(M) node[pos=0.5,right,above]{$\psi_{6tk}$};
\path[line](Phon5)--(WordT) node[pos=0.5,left,above]{$1-\psi_{6tk}$};
\path[line](Phon5)--(C) node[pos=0.5,right]{$\psi_{6tk}$};
\path[line](WordL)--(AN) node[pos=0.5,right]{$1-\psi_{8}$};
\path[line](WordL)--(U) node[pos=0.5,left,above]{$\psi_{8}$};
\path[line](WordT)--(N) node[pos=0.5,right]{$1-\psi_{7k}$};
\path[line](WordT)--(F) node[pos=0.5,left, above]{$\psi_{7k}$};
\end{tikzpicture}
}
\caption{\label{fig:irtmpt2}
Modified IRT-MPT model graph. Observable states have sharp corners and latent states have rounded corners.}
\end{figure}

Given the direct linking of the observable nodes to only some of the possible latent processes and the multiplicity of the linking of several outcomes to latent processes that are higher in the process hierarchy, it is reasonable to be worried about potential identifiability issues. Below we reproduce in Figure \ref{fig:irtmpt1} the original diagram of the IRT-MPT model, and in Figure \ref{fig:irtmpt2} we provide a modified version of the diagram (Figure \ref{fig:irtmpt2}) that better demonstrates the potential for identifiability issues. In Figure \ref{fig:irtmpt2} we collapse the multiplicities of the observable nodes that appear in Figure \ref{fig:irtmpt1} while conveying the same information as a directed acyclic graph (DAG). Instead of appearing as leaves in the tree as they are in Figure \ref{fig:irtmpt1}, the observable nodes are terminal nodes of paths in the DAG in Figure \ref{fig:irtmpt2}. This DAG has a unique root node (Attempt) and eight terminal nodes (the observable categories). Four observables (NA, S, M, C) have unique directed paths from the root, one observable (AN) has two length paths from the root , two observables (U, N) have three paths from the root, and one terminal node has a four paths from the root.

Identifiability and estimability in the context of this model need to be addressed individually. Identifiability is defined by different parameter settings leading to different distributions on the sample space of observable data. This is an inherent trait of the model and is tied to whether the parameters of the model could be perfectly reconstructed over an infinite number of independent and identical draws from an instance (particular choice of parameter values) of the process. Estimability is tied more closely to the observed data. The parameter is estimable from the observed data if derived estimates of the parameter are unique. This is also strongly connected to what is being estimated and what the estimation methodology is. In a purely likelihood context, this is usually associated with the uniqueness of the maximum likelihood estimator. In the context of Bayes estimators, prior assumptions often penalize all points on a likelihood ridge differently, which leads to unique estimators due solely to prior influence. Because of this, and the general arbitrariness of prior assumptions, it is often best to determine Bayes estimability in an objective prior framework (for instance under a Jeffreys' prior). However, this often reduces to the same problem one would be addressing when investigating estimability using likelihoods and maximum likelihood estimators.

\section{Model Architecture}

Figure \ref{fig:irtmpt2} shows the structure of the MPT model; there the probability of successfully performing each of the processes is denoted by the probabilities $\psi_{stk}$, where $s=1,2,\ldots,8$ indexes the cognitive process, $t$ the respondent and $k$ the word item. The probabilities for $s\in\{1, 3, 4, 5, 6\}$ (\emph{Attempt}, \emph{LexSem}, \emph{LexPhon}, \emph{LexSel}, and \emph{Phon}, respectively) are defined in terms of respondent skill $\theta_{ts}$ and item difficulties $\delta_{ks}$ by $\log (\psi_{stk} /(1 - \psi_{stk})) = \theta_{ts} -\delta_{ks}$. For $T$ respondents evaluated on a set of $K$ common items, this leads to a $6\times (T +K )+1$ dimensional parameter space (five $\theta_{ts}$ and one $\psi_{2t}$ per respondent, five $\delta_{ks}$ and one $\psi_{7k}$ per item, and one global $\psi_8$). Given the fact that the number of observed values, $T\times K$, is greater than the number of parameters, $6\times (T +K)+1$, this suggests an identifiable model so long as there are no structural identification issues in the model. 

One structural issue that is common in IRT models is that adding a constant to all $\theta_{ts}$ and $\delta_{ks}$ for a given process s leads to identical latent probabilities. This is usually ameliorated by forcing the average of either the respondent skills or the item difficulties to be 0 for each process $s$. The observed outcome is dependent on which processes the subject performs correctly, and corresponds to one of eight possible response categories: \emph{Correct} (C), \emph{Semantic} (S), \emph{Formal} (F), \emph{Mixed} (M), \emph{Unrelated} (U), \emph{Neologism} (N), \emph{Abstruse Neologism} (AN), and \emph{Non-naming Attempt} (NA). 
%
%
%

In the model, the respondent-item dependent probabilities $\psi_{stk}$ are represented in terms of the ability $\theta_{ts}$, item difficulty $\delta_{ks}$, and include an intercept term $\beta_s$, such that $\psi_{stk}$ is given by
\begin{equation}
\label{eq:psi_def_2}
\log\left(
\frac{\psi_{stk}}{1-\psi_{stk}} 
\right)
= \theta_{ts}-\delta_{ks}+\beta_s
\end{equation}
and specify two points in $\mathbb{R}^{T+K+1}$ to be equivalent if
\begin{equation}
\label{eq:psi_equivalence_relation_2}
\left(\boldsymbol{\theta}_{\cdot s}^\prime,\boldsymbol{\delta}_{\cdot s}^\prime,\beta_{s}^\prime\right)
-
\left(\boldsymbol{\theta}_{\cdot s},\boldsymbol{\delta}_{\cdot s},\beta_{s}\right)
= u_s(\boldsymbol{1}_{T},\boldsymbol{0}_K,0) + v_s(\boldsymbol{0}_{T},\boldsymbol{1}_K,0)+(v_s-u_s)(\boldsymbol{0}_{T},\boldsymbol{0}_K,1)
\end{equation}
for any $(u_s,v_s)\in\mathbb{R}^2$. Furthermore, we impose two linear restrictions on $\mathbb{R}^{T+K+1}$. The standard choice for representing the equivalence class is to make the restrictions
\begin{equation}
\label{eq:psi_linear_subspace_2}
\sum_{t=1}^T \theta_{ts}=0
\quad\text{and}\quad
\sum_{k=1}^K \delta_{ks}=0.
\end{equation}
for each $s\in\{1,3,\ldots,6\}$ we have a $T+K-1$ dimensional parameter. This provides $5(T+K-1)$ parameters for the $\psi$ that depend of $t$ and $k$. The total number of parameters is $\dim_{\model}=5(T+K-1)+T+K+1 = 6T+6K-4$ total parameters. 

\section{Basic probability equations to use}

\begin{subequations}
\begin{eqnarray}
p_{1tk} &=& P(R_{tk}\neq NA)\\
p_{2tk} &=& 
P(R_{tk}\notin\{AN,U,S\}|R_{tk}\neq NA)\\
p_{3tk} &=& 
P(R_{tk}=S|R_{tk}\in\{AN,U,S\})\\
p_{4tk} &=& 
P(R_{tk}=AN|R_{tk}\in\{AN,U,S\})
\\
p_{5tk} &=& P(R_{tk}=C|R_{tk}\notin\{NA,AN,U,S\}) 
\\
p_{6tk} &=& 
P(R_{tk}=M|R_{tk}\notin\{NA,AN,U,S\})
\\
p_{7tk} &=& 
P(R_{tk}=N|R_{tk}\notin\{NA,AN,U,S\})
\end{eqnarray}
\end{subequations}

\begin{subequations}
\begin{eqnarray}
\label{eq:probs_start_a}
p_{1tk} 
&=& \psi_{1tk}
\\
\label{eq:probs_start_b}
p_{2tk}
&=& \psi_{2t}\psi_{3tk}
\\
p_{3tk}
 &=&
\label{eq:probs_start_c}
\frac{\psi_{2t}(1-\psi_{3tk})\psi_{6tk}}{1-p_{2tk}}
\\
p_{4tk}
&=&
\label{eq:probs_start_d}
(1-\psi_{6tk})(1-\psi_8)
\\
p_{5tk}
\label{eq:probs_start_e}
&=&
\psi_{4tk}\psi_{5tk}\psi_{6tk}
\\
p_{6tk}
&=&
\label{eq:probs_start_f}
\psi_{4tk}(1-\psi_{5tk})\psi_{6tk}
\\
p_{7tk}
&=&
\label{eq:probs_start_g}
(1-\psi_{6tk})(1-\psi_{7k})
\end{eqnarray}
\end{subequations}

\begin{subequations}
\begin{eqnarray}
\label{eq:probs_sub_a}
\frac{p_{5tk}}{p_{6tk}} &=&
\frac{\psi_{5tk}}{1-\psi_{5tk}}\\
\label{eq:probs_sub_b}
p_{5tk}+p_{6tk}&=&
\psi_{4tk}\psi_{6tk}\\
\label{eq:probs_sub_c}
\frac{(1-p_{2tk})p_{3tk}}{p_{2tk}} &=&
\frac{(1-\psi_{3tk})\psi_{6tk}}{\psi_{3tk}}\\
\label{eq:probs_sub_d}
\frac{p_{7tk}}{p_{4tk}} &=&
\frac{1-\psi_{7k}}{1-\psi_8}
\end{eqnarray}
\end{subequations}

The equations that might even be better for going after the problem are \eqref{eq:probs_sub_a}-\eqref{eq:probs_sub_d} with \eqref{eq:probs_start_a}, \eqref{eq:probs_start_c}, and  \eqref{eq:probs_start_d}. If we have different parameter vectors $\boldsymbol{\omega}$ and $\boldsymbol{\omega}^\prime$ that lead to the same probabilities, then the left hand sides of these equations are the same. We get
\begin{subequations}
\begin{eqnarray}
\label{eq:equalities_a}
\psi_{1tk}
&=& \psi_{1tk}^\prime\\
\label{eq:equalities_b}
\frac{\psi_{5tk}}{1-\psi_{5tk}}&=&
\frac{\psi_{5tk}^\prime}{1-\psi_{5tk}^\prime}\\
\label{eq:equalities_c}
\psi_{4tk}\psi_{6tk}&=&
\psi_{4tk}^\prime\psi_{6tk}^\prime\\
\label{eq:equalities_d}
\psi_{2t}\psi_{3tk}&=&
\psi_{2t}^\prime\psi_{3tk}^\prime\\
\label{eq:equalities_e}
\frac{(1-\psi_{3tk})\psi_{6tk}}{\psi_{3tk}}&=&
\frac{\left(1-\psi_{3tk}^\prime\right)\psi_{6tk}^\prime}{\psi_{3tk}^\prime}\\
\label{eq:equalities_f}
(1-\psi_{6tk})(1-\psi_8)
&=&
\left(1-\psi_{6tk}^\prime\right)\left(1-\psi_8^\prime\right)
\\
\label{eq:equalities_g}
\frac{1-\psi_{7k}}{1-\psi_8} &=&
\frac{1-\psi_{7k}^\prime}{1-\psi_8^\prime}
\end{eqnarray}
\end{subequations}

\section{Identifiability analysis at a high level}
The first thing to notice is that the mapping
$
\left(\boldsymbol{\theta}_{\cdot s},\boldsymbol{\delta}_{\cdot s},\beta_s\right)
\mapsto \boldsymbol{\psi}_{s\cdot\cdot}
$
is injective. So if we get equalities of the $\boldsymbol{\psi}_{s\cdot\cdot}$ and $\boldsymbol{\psi}_{s\cdot\cdot}^\prime$, then we get equalities of
$\left(\boldsymbol{\theta}_{\cdot s},\boldsymbol{\delta}_{\cdot s},\beta_s\right)$ and $\left(\boldsymbol{\theta}_{\cdot s}^\prime,\boldsymbol{\delta}_{\cdot s}^\prime,\beta_s^\prime\right)$.\\
From \eqref{eq:equalities_a}, we get  $\psi_{1tk}^\prime=\psi_{1tk}$ for all $(t,k)$ and so $\left(\boldsymbol{\theta}_{\cdot1}^\prime,\boldsymbol{\delta}_{\cdot1}^\prime,\beta_1^\prime\right)=\left(\boldsymbol{\theta}_{\cdot1},\boldsymbol{\delta}_{\cdot1},\beta_1\right)$.
\\
From \eqref{eq:equalities_b}, we get $\psi_{5tk}^\prime=\psi_{5tk}$ for all $(t,k)$ and so $\left(\boldsymbol{\theta}_{\cdot5}^\prime,\boldsymbol{\delta}_{\cdot5}^\prime,\beta_5^\prime\right)=\left(\boldsymbol{\theta}_{\cdot5},\boldsymbol{\delta}_{\cdot5},\beta_5\right)$.
\\
Define $\eta=\frac{1-\psi_8}{1-\psi_8^\prime}$ so that $\psi_8^\prime = 1-\frac{1}{\eta}+\frac{\psi_8}{\eta}$ and so $\eta>1-\psi_8$.
\\
From \eqref{eq:equalities_g}, we get that 
\[
\eta=\frac{1-\psi_{7k}}{1-\psi_{7k}^\prime}
\]
for all $k$.
We get
\[
\psi_{7k}^\prime
=
1-\frac{1}{\eta}+\frac{\psi_{7k}}{\eta}
\]
Another restriction on $\eta$ that is necessary is $\eta>1-\psi_{7k}$ for all $k$.
\\
From \eqref{eq:equalities_f}, we get that 
\begin{equation}
\label{eq:eta_psi_6_eq}
\eta=\frac{1-\psi_{6tk}^\prime}{1-\psi_{6tk}}
\end{equation}
for all $(t,k)$. Another restriction on $\eta$ that is $0<\eta<\frac{1}{1-\psi_{6tk}}$ for all $(t,k)$. 
\\ 
The last equality is equivalent to
\[
\psi_{6tk}^\prime = 1-\left(1-\psi_{6tk}\right)\eta = 1-\eta+\eta\psi_{6tk}
\]
We plug this into \eqref{eq:equalities_e} and get
\begin{equation}
\label{eq:eta_psi_3_eq}
\frac{(1-\psi_{3tk})\psi_{6tk}}{\psi_{3tk}}=
\frac{\left(1-\psi_{3tk}^\prime\right)\psi_{6tk}^\prime}{\psi_{3tk}^\prime}
=
\frac{\left(1-\psi_{3tk}^\prime\right)\left(1-\eta+\eta\psi_{6tk}\right)}{\psi_{3tk}^\prime}
\end{equation}
This tells us that
\[
\begin{array}{rcl}
\psi_{3tk}^\prime
&=&
\left(1+
\frac{(1-\psi_{3tk})\psi_{6tk}}{\psi_{3tk}\left(1-\eta+\eta\psi_{6tk}\right)}
\right)^{-1}
\\&=&
\frac{\psi_{3tk}\left(1-\eta+\eta\psi_{6tk}\right)}{\psi_{3tk}\left(1-\eta+\eta\psi_{6tk}\right)+(1-\psi_{3tk})\psi_{6tk}}
\\&=&
\frac{\psi_{3tk}\left(1-\eta+\eta\psi_{6tk}\right)}{\psi_{3tk}\left(1-\eta\right)\left(1-\psi_{6tk}\right)+\psi_{6tk}}
\end{array}
\]
This $\psi_{3tk}$ is always between zero and one, and so no new restriction on $\eta$ is needed.
\\
Plugging these into \eqref{eq:equalities_d}, we get
\begin{equation}
\label{eq:eta_psi_2_eq}
\begin{array}{rcl}
\psi_{2t}^\prime
&=&
\frac{\psi_{2t}\psi_{3tk}}{\psi_{3tk}^\prime}
\\&=&
\psi_{2t}\psi_{3tk}
\frac{\psi_{3tk}\left(1-\eta+\eta\psi_{6tk}\right)+(1-\psi_{3tk})\psi_{6tk}}{\psi_{3tk}\left(1-\eta+\eta\psi_{6tk}\right)}
\\&=&
\psi_{2t}\psi_{3tk}
+
\psi_{2t}(1-\psi_{3tk})\frac{\psi_{6tk}}{1-\eta+\eta\psi_{6tk}}
\end{array}
\end{equation}
We get a further restriction on $\eta$
\[
\begin{array}{rcl}
1&>&
\psi_{2t}\psi_{3tk}
+
\psi_{2t}(1-\psi_{3tk})\frac{\psi_{6tk}}{1-\eta+\eta\psi_{6tk}}\\
\frac{1-\psi_{2t}\psi_{3tk}}{\psi_{2t}(1-\psi_{3tk})\psi_{6tk}}&>&
\frac{1}{1-\eta+\eta\psi_{6tk}}\\
\frac{\psi_{2t}(1-\psi_{3tk})\psi_{6tk}}{1-\psi_{2t}\psi_{3tk}}&<&
1-\eta+\eta\psi_{6tk}\\
\eta(1-\psi_{6tk})&<&
1-\frac{\psi_{2t}(1-\psi_{3tk})\psi_{6tk}}{1-\psi_{2t}\psi_{3tk}}
\\
\eta &<&
\frac{1-\frac{\psi_{2t}(1-\psi_{3tk})\psi_{6tk}}{1-\psi_{2t}\psi_{3tk}}}{1-\psi_{6tk}}
\\
\eta &<&
\frac{1-\psi_{2t}\psi_{3tk}-\psi_{2t}(1-\psi_{3tk})\psi_{6tk}}{(1-\psi_{2t}\psi_{3tk})(1-\psi_{6tk})}
\\
\eta &<&
\frac{1-\psi_{2t}(\psi_{3tk}+(1-\psi_{3tk})\psi_{6tk})}{(1-\psi_{2t}\psi_{3tk})(1-\psi_{6tk})}
\end{array}
\]
\\
Using \eqref{eq:equalities_c}, we get
\begin{equation}
\label{eq:eta_psi_4_eq}
\psi_{4tk}^\prime
=
\frac{\psi_{4tk}\psi_{6tk}}{1-\eta+\eta\psi_{6tk}}
\end{equation}
We get one final restriction on $\eta$
\[
\eta<\frac{\left(1-\psi_{4tk}\psi_{6tk}\right)}{\left(1-\psi_{6tk}\right)}
\]
\\
We get a range of values for $\boldsymbol{\omega}$ that could possibly admit a transformation
\[
\max\left\{1-\psi_8,\max_k\{1-\psi_{7k}\}\right\}
< \eta
< \min_{t,k}\left\{\frac{
1-\psi_{6tk}
\max \left\{
\psi_{4tk},
\frac{\psi_{2t}-\psi_{2t}\psi_{3tk}}{1-\psi_{2t}\psi_{3tk}}
\right\}
}{1-\psi_{6tk}}
\right\}
\]

\section{The details of parameter restrictions}

From \eqref{eq:equalities_f}, we get that 
\begin{equation}
\label{eq:eta_psi_6_eq_second}
\eta=\frac{1-\psi_{6tk}^\prime}{1-\psi_{6tk}}
\end{equation}
for all $(t,k)$.
\\
We want to see if we can get a contradiction of the existence of this mapping or get a restriction on the values of the parameter that can admit such a mapping. 
The former will provide identifiability 
and the latter will provide identifiability 
for parameter values in a set of full Lebesgue measure.
\\
\eqref{eq:eta_psi_6_eq_second} is equivalent to
\[
\frac{\psi_{6tk}^\prime}{1-\psi_{6tk}^\prime} 
= 
\frac{1-\left(1-\psi_{6tk}\right)\eta}{\left(1-\psi_{6tk}\right)\eta}
=
\frac{\frac{1}{\left(1-\psi_{6tk}\right)}-\eta}{\eta}
=
\frac{1+\frac{\psi_{6tk}}{\left(1-\psi_{6tk}\right)}-\eta}{\eta}
\]
and so
\begin{equation}
\label{eq:eta_psi_6_eq_theta_delta_beta}
\exp\left(\theta_{t 6}^\prime-\delta_{k6}^\prime+\beta_6^\prime\right) 
= 
\frac{1+\exp\left(\theta_{t 6}-\delta_{k6}+\beta_6\right) -\eta}{\eta}
\end{equation}
For different respondents $t_1$ and $t_2$ doing the same item $k$ this means that
\begin{equation}
\label{eq:eta_psi_6_eq_same_item}
\exp\left(\theta_{t_1 6}^\prime-\theta_{t_2 6}^\prime\right)
=
\frac{1+\exp\left(\theta_{t_1 6}-\delta_{k6}+\beta_6\right) -\eta}{1+\exp\left(\theta_{t_2 6}-\delta_{k6}+\beta_6\right) -\eta}
\end{equation}
where the right hand side cannot depend on $k$ because the left hand side does not depend on $k$.
\\
For the same respondent on different items, we have
\begin{equation}
\label{eq:eta_psi_6_eq_same_respondent}
\exp\left(\delta_{k_26}^\prime-\delta_{k_16}^\prime\right) 
= 
\frac{1+\exp\left(\theta_{t 6}-\delta_{k_16}+\beta_6\right) -\eta}{1+\exp\left(\theta_{t 6}-\delta_{k_26}+\beta_6\right) -\eta}
\end{equation}
where the right hand side cannot depend on $t$ because the left hand side does not depend on $t$.
\\
What are the implications of \eqref{eq:eta_psi_6_eq_same_item} and \eqref{eq:eta_psi_6_eq_same_respondent}?\\
First, one way for \eqref{eq:eta_psi_6_eq_same_item} to hold for all $k$ is for $\boldsymbol{\theta}_{\cdot 6}=\boldsymbol{\theta}_{\cdot 6}^\prime=\boldsymbol{0}$ and \eqref{eq:eta_psi_6_eq_same_item} reduces to $1=1$. This means that in
\eqref{eq:eta_psi_6_eq_theta_delta_beta} we have
\begin{equation}
\label{eq:eta_psi_6_eq_theta0}
\exp\left(-\delta_{k6}^\prime+\beta_6^\prime\right) 
= 
\frac{1+\exp\left(-\delta_{k6}+\beta_6\right) -\eta}{\eta}
\end{equation}
We must have
\[
\exp\left(K\beta_6^\prime\right) 
= 
\prod_{j=1}^K
\frac{1+\exp\left(-\delta_{j6}+\beta_6\right) -\eta}{\eta}
\]
and
\[
\exp\left(-\delta_{k6}^\prime\right)
=
\frac{\frac{1+\exp\left(-\delta_{k6}+\beta_6\right) -\eta}{\eta}}{
\left(\prod_{j=1}^K
\frac{1+\exp\left(-\delta_{j6}+\beta_6\right) -\eta}{\eta}
\right)^{\frac{1}{K}}
}
=
\frac{1+\exp\left(-\delta_{k6}+\beta_6\right) -\eta}{
\left(\prod_{j=1}^K
1+\exp\left(-\delta_{j6}+\beta_6\right) -\eta
\right)^{\frac{1}{K}}
}
\]
These definitions provide \eqref{eq:eta_psi_6_eq_same_respondent} in this case and so we apparently have a possible mapping.
\\
Second, if we do not assume that $\boldsymbol{\theta}_{\cdot 6}=\boldsymbol{\theta}_{\cdot 6}^\prime=\boldsymbol{0}$, then we can instead assume that $\boldsymbol{\delta}_{\cdot 6}=\boldsymbol{\delta}_{\cdot 6}^\prime=\boldsymbol{0}$. This provides \eqref{eq:eta_psi_6_eq_same_respondent} and 
\begin{equation}
\label{eq:eta_psi_6_eq_delta0}
\exp\left(\theta_{k6}^\prime+\beta_6^\prime\right) 
= 
\frac{1+\exp\left(\theta_{k6}+\beta_6\right) -\eta}{\eta}
\end{equation}
from \eqref{eq:eta_psi_6_eq_theta_delta_beta}. We must have
\[
\exp\left(T\beta_6^\prime\right) 
= 
\prod_{i=1}^T
\frac{1+\exp\left(\theta_{i6}+\beta_6\right) -\eta}{\eta}
\]
and
\[
\exp\left(\theta_{t6}^\prime\right)
=
\frac{\frac{1+\exp\left(\theta_{t6}+\beta_6\right) -\eta}{\eta}}{
\left(\prod_{i=1}^T
\frac{1+\exp\left(\theta_{i6}+\beta_6\right) -\eta}{\eta}
\right)^{\frac{1}{T}}
}
=
\frac{1+\exp\left(\theta_{t6}+\beta_6\right) -\eta}{
\left(\prod_{i=1}^T
1+\exp\left(-\theta_{i6}+\beta_6\right) -\eta
\right)^{\frac{1}{T}}
}
\]
These definitions provide \eqref{eq:eta_psi_6_eq_same_item} in this case and so we apparently have a possible mapping.
\\
Third, let us assume that neither of the above cases hold. Then we should be able to reach a contradiction.
Let $t_1$ and $t_2$ be such that $\theta_{t_16}\neq\theta_{t_26}$ and $k_1$ and $k_2$ be such that $\delta_{k_16}\neq\delta_{k_26}$. Then we must have
\[
\frac{1+\exp\left(\theta_{t_1 6}-\delta_{k_16}+\beta_6\right) -\eta}{1+\exp\left(\theta_{t_1 6}-\delta_{k_26}+\beta_6\right) -\eta}
=
\frac{1+\exp\left(\theta_{t_2 6}-\delta_{k_16}+\beta_6\right) -\eta}{1+\exp\left(\theta_{t_2 6}-\delta_{k_26}+\beta_6\right) -\eta}
\]
and so
\[
\begin{array}{rl}
&
\left(
1+\exp\left(\theta_{t_1 6}-\delta_{k_16}+\beta_6\right) -\eta
\right)
\left(
1+\exp\left(\theta_{t_2 6}-\delta_{k_26}+\beta_6\right) -\eta
\right)
\\=&
\left(
1+\exp\left(\theta_{t_2 6}-\delta_{k_16}+\beta_6\right) -\eta
\right)
\left(
1+\exp\left(\theta_{t_1 6}-\delta_{k_26}+\beta_6\right) -\eta
\right)
\\&
\left(
\exp\left(\theta_{t_1 6}-\delta_{k_16}\right)
+
\exp\left(\theta_{t_2 6}-\delta_{k_26}\right)
\right)
\\=&
\left(
\exp\left(\theta_{t_1 6}-\delta_{k_26}\right)
+
\exp\left(\theta_{t_2 6}-\delta_{k_16}\right)
\right)
\\&
\left(
\exp\left(\theta_{t_1 6}-\theta_{t_2 6}+\delta_{k_26}-\delta_{k_16}\right)
+
1
\right)
\\=&
\left(
\exp\left(\theta_{t_1 6}-\theta_{t_2 6}\right)
+
\exp\left(\delta_{k_26}-\delta_{k_16}\right)
\right)
\end{array}
\]
which is an equation of the form 
\[
\exp(x)\exp(y) +1 = \exp(x)+\exp(y)
\]
isolating $\exp(x)$ we get
\[
\exp(x)(\exp(y)-1) = \exp(y)-1
\]
and so $y\neq0$ requires $x=0$ and $x\neq0$ requires $y=0$. Because $(t_1,t_2)$ and $(k_1,k_2)$ were arbitrary, we must be in one of the two cases already discussed.

\subsection{Assuming $\boldsymbol{\theta}_{\cdot 6}=\boldsymbol{0}$ and  $\boldsymbol{\delta}_{\cdot 6}\neq\boldsymbol{0}$}
We work through implications for other part of the parameter space from the assumption of this subcase and the existence of an identifiability issue.
For this analysis, we will let $\psi_{6tk}=\psi_{6k}$ and $\psi_{6tk}^\prime=\psi_{6k}^\prime$ because neither can depend on $t$.

\subsubsection{Implications for $\psi_{3tk}$}
From \eqref{eq:equalities_e}, we have
\[
\exp\left(
\theta_{t3}^\prime-\theta_{t3}+\delta_{k3}-\delta_{k3}^\prime+\beta_3^\prime-\beta_3
\right)
=
\frac{\psi_{6k}^\prime}{\psi_{6k}}
\]
For different values $t_1$ and $t_2$, we have to have
\[
\exp\left(
\theta_{t_13}^\prime-\theta_{t_13}
\right)
=
\exp\left(
\theta_{t_23}^\prime-\theta_{t_23}
\right)
\]
So we have
\[
\exp\left(
\theta_{t_13}^\prime-\theta_{t_23}^\prime
\right)
=
\exp\left(
\theta_{t_13}-\theta_{t_23}
\right)
\]
Taking the geometric average over $t_2$, we get
\[
\exp\left(
\theta_{t_13}^\prime
\right)
=
\exp\left(
\theta_{t_13}
\right)
\]
and so $\theta_{t3}^\prime=\theta_{t3}$ for all $t$.

Now, we have
\[
\exp\left(
\delta_{k3}-\delta_{k3}^\prime+\beta_3^\prime-\beta_3
\right)
=
\frac{\psi_{6k}^\prime}{\psi_{6k}}
\]
and taking the geometric average over $k$ provides
\[
\exp\left(
\beta_3^\prime
\right)
=
\exp\left(
\beta_3
\right)
\left(
\prod_{k=1}^K
\frac{\psi_{6k}^\prime}{\psi_{6k}}
\right)^{\frac{1}{K}}
=
\exp\left(
\beta_3
\right)
\left(
\prod_{k=1}^K
\frac{1-\eta+\eta\psi_{6k}}{\psi_{6k}}
\right)^{\frac{1}{K}}
\]
and so we also have
\[
\exp\left(
-\delta_{k3}^\prime
\right)
=
\exp\left(
-\delta_{k3}-\beta_3^\prime+\beta_3
\right)
\frac{\psi_{6k}^\prime}{\psi_{6k}}
=
\exp\left(
-\delta_{k3}
\right)
\frac{1-\eta+\eta\psi_{6k}}{\psi_{6k}}
\left(
\prod_{\tilde{k}=1}^K
\frac{\psi_{6\tilde{k}}}{1-\eta+\eta\psi_{6\tilde{k}}}
\right)^{\frac{1}{K}}
\]

Finally, we get
\[
\exp(\theta_{t3}^\prime-\delta_{k3}^\prime+\beta_3^\prime)
=
\exp(\theta_{t3}-\delta_{k3}+\beta_3)\frac{1-\eta+\eta\psi_{6k}}{\psi_{6k}}
\]
and so, as expected, we get
\[
\psi_{3tk}^\prime = 
\frac{\exp(\theta_{t3}-\delta_{k3}+\beta_3)\frac{1-\eta+\eta\psi_{6k}}{\psi_{6k}}}{1+\exp(\theta_{t3}-\delta_{k3}+\beta_3)\frac{1-\eta+\eta\psi_{6k}}{\psi_{6k}}}
=
\frac{\psi_{3tk}(1-\eta+\eta\psi_{6k})}{(1-\psi_{3tk})\psi_{6k}+\psi_{3tk}(1-\eta+\eta\psi_{6k})}
\]
and we have described the mapping on the parameter space that allows this to happen. Notice that the mapping is all about fixing the parameter that changes with $t$ and allows the parameters that depend on $k$ to change in a way that is consistent with the $\psi_{6k}\mapsto\psi_{6k}^\prime$ mapping.

\subsubsection{Implications for $\psi_{4tk}$}
From \eqref{eq:equalities_c}, we have
\[
\psi_{4tk}^\prime=
\psi_{4tk}\frac{\psi_{6k}}{1-\eta+\eta\psi_{6k}}
\]
which is
\[
\frac{\exp(\theta_{t4}^\prime-\delta_{k4}^\prime+\beta_4^\prime)}{1+\exp(\theta_{t4}^\prime-\delta_{k4}^\prime+\beta_4^\prime)}
=
\frac{\exp(\theta_{t4}-\delta_{k4}+\beta_4)}{1+\exp(\theta_{t4}-\delta_{k4}+\beta_4)}
\frac{\psi_{6k}}{1-\eta+\eta\psi_{6k}}
\]
and so
\[
\frac{\exp(\theta_{t4}^\prime-\delta_{k4}^\prime+\beta_4^\prime)}{1+\exp(\theta_{t4}^\prime-\delta_{k4}^\prime+\beta_4^\prime)}
\frac{1+\exp(\theta_{t4}-\delta_{k4}+\beta_4)}{\exp(\theta_{t4}-\delta_{k4}+\beta_4)}
=
\frac{\psi_{6k}}{1-\eta+\eta\psi_{6k}}
\]
depends only on $k$. Rewriting a bit we get
\[
\frac{\exp(-\theta_{t4})+\exp(-\delta_{k4}+\beta_4)}{\exp(-\theta_{t4}^\prime)+\exp(-\delta_{k4}^\prime+\beta_4^\prime)}
\frac{\exp(-\delta_{k4}^\prime+\beta_4^\prime)}{\exp(-\delta_{k4}+\beta_4)}
=
\frac{\psi_{6k}}{1-\eta+\eta\psi_{6k}}
\]
and the right hand side can't depend on $t$ and so we need $\boldsymbol{\theta}_{\cdot 4} = \boldsymbol{\theta}_{\cdot 4}^\prime = \boldsymbol{0}$. We get
\[
\frac{1+\exp(-\delta_{k4}+\beta_4)}{1+\exp(-\delta_{k4}^\prime+\beta_4^\prime)}
\frac{\exp(-\delta_{k4}^\prime+\beta_4^\prime)}{\exp(-\delta_{k4}+\beta_4)}
=
\frac{\psi_{6k}}{1-\eta+\eta\psi_{6k}}
\]
and we get
\[
\frac{1}{1+\exp(\delta_{k4}^\prime-\beta_4^\prime)}
=
\frac{\exp(-\delta_{k4}+\beta_4)}{1+\exp(-\delta_{k4}+\beta_4)}\frac{\psi_{6k}}{1-\eta+\eta\psi_{6k}}
\]
and
\[
\exp(\delta_{k4}^\prime-\beta_4^\prime)
=
\left(1+\exp(\delta_{k4}-\beta_4)\right)\frac{1-\eta+\eta\psi_{6k}}{\psi_{6k}}
-1
\]
Taking geometric average over $k$, we get
\[
\exp(-\beta_4^\prime)
=
\left(\prod_{k=1}^K
\left(\left(1+\exp(\delta_{k4}-\beta_4)\right)\frac{1-\eta+\eta\psi_{6k}}{\psi_{6k}}
-1\right)\right)^{\frac{1}{K}}
\]
and so we also get
\[
\exp(\delta_{k4}^\prime)
=
\frac{\left(\left(1+\exp(\delta_{k4}-\beta_4)\right)\frac{1-\eta+\eta\psi_{6k}}{\psi_{6k}}
-1\right)}{
\left(\prod_{k=1}^K
\left(\left(1+\exp(\delta_{k4}-\beta_4)\right)\frac{1-\eta+\eta\psi_{6k}}{\psi_{6k}}
-1\right)\right)^{\frac{1}{K}}}
\]
Of course, we could have just noted that $\psi_{4tk}=\psi_{4k}$ and $\psi_{4tk}^\prime=\psi_{4k}^\prime$ and get
\[
\psi_{4k}^\prime = 
\psi_{4k}\frac{\psi_{6k}}{1-\eta+\eta\psi_{6k}}
\]

\subsubsection{Implications for $\psi_{2t}$}
From \eqref{eq:equalities_d} we get
\[
\frac{\psi_{2t}}{\psi_{2t}^\prime}
=
\frac{1-\eta+\eta\psi_{6k}}{(1-\psi_{3tk})\psi_{6k}+\psi_{3tk}(1-\eta+\eta\psi_{6k})}
=
\frac{1}{1-\psi_{3tk}}
\frac{1}{
\frac{\psi_{6k}}{1-\eta+\eta\psi_{6k}}
+
\frac{\psi_{3tk}}{1-\psi_{3tk}}
}
\]
which is
\[
\frac{\psi_{2t}}{\psi_{2t}^\prime}
=
\frac{1+\exp(\theta_{t3}-\delta_{k3}+\beta_3)}{
\frac{\psi_{6k}}{1-\eta+\eta\psi_{6k}}
+
\exp(\theta_{t3}-\delta_{k3}+\beta_3)
}
=
\frac{\exp(\delta_{k3}-\beta_3)+\exp(\theta_{t3})}{
\frac{\psi_{6k}\exp(\delta_{k3}-\beta_3)}{1-\eta+\eta\psi_{6k}}
+
\exp(\theta_{t3})
}
\]
The left hand side does not depend on $k$, and so for $k_1$ and $k_2$ we have
\[
\frac{\exp(\delta_{k_13}-\beta_3)+\exp(\theta_{t3})}{
\frac{\psi_{6k_1}\exp(\delta_{k_13}-\beta_3)}{1-\eta+\eta\psi_{6k_1}}
+
\exp(\theta_{t3})
}
=
\frac{\exp(\delta_{k_23}-\beta_3)+\exp(\theta_{t3})}{
\frac{\psi_{6k_2}\exp(\delta_{k_23}-\beta_3)}{1-\eta+\eta\psi_{6k_2}}
+
\exp(\theta_{t3})
}
\]
and this gives
\[
\begin{array}{rl}
&
\frac{\psi_{6k_2}\exp(\delta_{k_23}-\beta_3)}{1-\eta+\eta\psi_{6k_2}}\exp(\delta_{k_13}-\beta_3)
-
\frac{\psi_{6k_1}\exp(\delta_{k_13}-\beta_3)}{1-\eta+\eta\psi_{6k_1}}\exp(\delta_{k_23}-\beta_3)
\\&=
\exp(\theta_{t3})
\left(
\frac{\psi_{6k_1}\exp(\delta_{k_13}-\beta_3)}{1-\eta+\eta\psi_{6k_1}}+\exp(\delta_{k_23}-\beta_3)
-
\frac{\psi_{6k_2}\exp(\delta_{k_23}-\beta_3)}{1-\eta+\eta\psi_{6k_2}}-\exp(\delta_{k_13}-\beta_3)
\right)
\end{array}
\]
and so for any $t$ and any $k_1$ and $k_2$ we must have
\[
\exp(\theta_{t3})
=
\frac{
\frac{\psi_{6k_2}\exp(\delta_{k_23}-\beta_3)}{1-\eta+\eta\psi_{6k_2}}\exp(\delta_{k_13}-\beta_3)
-
\frac{\psi_{6k_1}\exp(\delta_{k_13}-\beta_3)}{1-\eta+\eta\psi_{6k_1}}\exp(\delta_{k_23}-\beta_3)
}{
\frac{\psi_{6k_1}\exp(\delta_{k_13}-\beta_3)}{1-\eta+\eta\psi_{6k_1}}+\exp(\delta_{k_23}-\beta_3)
-
\frac{\psi_{6k_2}\exp(\delta_{k_23}-\beta_3)}{1-\eta+\eta\psi_{6k_2}}-\exp(\delta_{k_13}-\beta_3)
}
\]
We get that $\boldsymbol{\theta}_{\cdot 3}=\boldsymbol{\theta}_{\cdot 3}^\prime=\boldsymbol{0}$ and we could just an easily call $\psi_{3tk}=\psi_{3k}$ and $\psi_{3tk}^\prime=\psi_{3k}^\prime$. The implication is that
\[
\frac{\psi_{2t}}{\psi_{2t}^\prime}
=
\frac{1-\eta+\eta\psi_{6k}}{(1-\psi_{3k})\psi_{6k}+\psi_{3k}(1-\eta+\eta\psi_{6k})}
\]
Because the left hand side does not depend on $t$ and the right hand side does not depend on $k$, there must be a constant $\xi$ such that
\[
\xi=
\frac{\psi_{2t}}{\psi_{2t}^\prime}
=
\frac{1-\eta+\eta\psi_{6k}}{(1-\psi_{3k})\psi_{6k}+\psi_{3k}(1-\eta+\eta\psi_{6k})}
\]
for all $t$ and $k$. Now, this makes a restriction on $\psi_{3k}$ in terms of $\xi$, $\psi_{6k}$, and $\eta$.
\[
\xi = 
\frac{1-\eta+\eta\psi_{6k}}{\psi_{6k}+\psi_{3k}(1-\eta)(1-\psi_{6k})}
\]
and so
\[
\psi_{3k}
=
\frac{
\frac{1-\eta+\eta\psi_{6k}}{\xi}-\psi_{6k}
}{
(1-\eta)(1-\psi_{6k})
}
\]
and of course
\[
\psi_{3k}^\prime = 
\psi_{3k}\xi
\]
Now, we need
\[
0<\frac{1-\eta+\eta\psi_{6k}}{\xi}-\psi_{6k}
\]
which gives
\[
\xi<\frac{1-\eta+\eta\psi_{6k}}{\psi_{6k}}
\]
We also need
\[
\frac{1-\eta+\eta\psi_{6k}}{\xi}-\psi_{6k}<
(1-\eta)(1-\psi_{6k})
\]
and so
\[
\xi>\frac{1-\eta+\eta\psi_{6k}}{(1-\eta)(1-\psi_{6k})+\psi_6} = 1
\]

\subsubsection{Pulling this all together}
If an identifiability issue exists, then we must have room for some $\eta$, which means
\[
\max\left\{1-\psi_8,\max_k\{1-\psi_{7k}\}\right\}
< \min_{t,k}\left\{\frac{
1-\psi_{6k}
\max \left\{
\psi_{4k},
\frac{\psi_{2t}-\psi_{2t}\psi_{3k}}{1-\psi_{2t}\psi_{3k}}
\right\}
}{1-\psi_{6k}}
\right\}
\]
There must be some $\xi$ that provides
\[
\psi_{3k}
=
\frac{
\frac{1-\eta+\eta\psi_{6k}}{\xi}-\psi_{6k}
}{
(1-\eta)(1-\psi_{6k})
}
\]

The way to generate such a parameter vector is to generate $\psi_8$, $\boldsymbol{\psi}_{7\cdot}$, $\boldsymbol{\psi}_{6\cdot}$, and $\boldsymbol{\psi}_{4\cdot}$ that depend only on $k$; $\boldsymbol{\psi}_{2\cdot}$ that depends only on $t$; $\boldsymbol{\psi}_{1\cdot\cdot}$ and $\boldsymbol{\psi}_{5\cdot\cdot}$ that depend on $t$ and $k$; generate a $\eta$ in the range
\[
\left(
\max\left\{1-\psi_8,\max_k\{1-\psi_{7k}\}\right\}
,
\min_{t,k}\left\{\frac{
1-\psi_{6k}
\max \left\{
\psi_{4k},
\psi_{2t}
\right\}
}{1-\psi_{6k}}
\right\}
\right)
\]
where we have taken $\psi_{3k}=1$ in the range upper bound calculation so that we can define $\psi_{3k}=1$ in terms of $\psi_{6k}$, $\eta$, and $\xi$; generate $\xi$ in the range
\[
\left(1,
\frac{1-\eta+\eta\psi_{6k}}{\psi_{6k}}
\right);
\]
and finally computing 
\[
\psi_{3k}
=
\frac{
\frac{1-\eta+\eta\psi_{6k}}{\xi}-\psi_{6k}
}{
(1-\eta)(1-\psi_{6k})
}
\]
As such, this indicates that the model is not identifiable.  Although there exist two additional cases, namely when $\boldsymbol{\delta}_{\cdot 6}=\boldsymbol{0}$ and $\boldsymbol{\theta}_{\cdot 6}\neq\boldsymbol{0}$ and when $\boldsymbol{\theta}_{\cdot 6}=\boldsymbol{0}$ and $\boldsymbol{\delta}_{\cdot 6}=\boldsymbol{0}$, their analysis is analogous to that of the case presented above, and it is therefore omitted for succinctness from this report.

\section*{Acknowledgements}

AW, DTR, GF, and WH were partially supported by NIH award 5R01DC018813-04. DTR was also partially supported by NSF RTG DMS award 2136228.

\bibliography{bibid}

\begin{thebibliography}{}

\bibitem[Fergadiotis et~al., 2015]{irt2015}
Fergadiotis, G., Kellough, S., and Hula, W.~D. (2015).
\newblock Item response theory modeling of the philadelphia naming test.
\newblock {\em Journal of Speech, Language, and Hearing Research},
  58(3):865--877.

\bibitem[Walker et~al., 2018]{walker}
Walker, G.~M., Hickok, G., and Fridriksson, J. (2018).
\newblock A cognitive psychometric model for assessment of picture naming
  abilities in aphasia.
\newblock {\em Psychological assessment}, 30(6):809.

\end{thebibliography}

\end{document}